# Sparse Aperture Masking (SAM) at NAOS/CONICA on the VLT

Peter Tuthill[a], Sylvestre Lacour[b], Paola Amico[c], Michael Ireland[a], Barnaby Norris[a], Paul Stewart[a], Tom Evans[a], Adam Kraus[d], Chris Lidman[e], Emanuela Pompei[c], Nicholas Kornweibel[c]

[a]School of Physics, University of Sydney, NSW 2006, Australia;
[b]Observatoire de Paris, LESIA/CNRS UMR XXX, 5 place Jules Janssen, Meudon, France
[c]European Southern Observatory, Karl-Schwarzschild-Strasse 2, 85748 Garching, Germany
[d]Institute for Astronomy, 2680 Woodlawn Drive, Honolulu, Hawaii 96822-1839, USA
[e]AAO, PO Box 296, Epping NSW 1710, Australia

## ABSTRACT

The new operational mode of aperture masking interferometry has been added to the CONICA camera which lies downstream of the Adaptive Optics (AO) corrected focus provided by NAOS on the VLT-UT4 telescope. Masking has been shown to deliver superior PSF calibration, rejection of atmospheric noise and robust recovery of phase information through the use of closure phases. Over the resolution range from about half to several resolution elements, masking interferometry is presently unsurpassed in delivering high fidelity imaging and direct detection of faint companions. Here we present results from commissioning data using this powerful new operational mode, and discuss the utility for masking in a variety of scientific contexts. Of particular interest is the combination of the CONICA polarimetry capabilities together with SAM mode operation, which has revealed structures never seen before in the immediate circumstellar environments of dusty evolved stars.

## 1. INTRODUCTION AND PRINCIPLES

The non-redundant masking technique[1] transforms the telescope into a sparse, separated-element array with a Fizeau-type beam combination scheme by the simple action of placing an apodizing plate over the pupil. It is instructive to examine the continuing success of this technique, which essentially dates back to Fizeau himself, in delivering unsurpassed image quality. When two or more pairs of coherent sub-apertures contribute to the same spatial frequency, that is they have the same baseline, then in a seeing-limited case, the power will add incoherently. The resultant baseline power will then be a random walk of $R$ steps where $R$ is the *redundancy* – the number of times the given baseline is repeated within the pupil. In the bright source limit, the noise inherent to this random walk process completely dominates the signal-to-noise (SNR), leading to the well-known result that the SNR of speckle frames saturates to 1. The success of masking can then be easily visualized: in order to create a non-redundant array one must discard most of the pupil area, and so lose most of the signal, but one also removes almost all of the atmospheric noise, and this leads to dramatically enhanced SNR.

The renaissance of aperture masking in the modern era can be traced back to John Baldwin's testbed demonstrating closure phase recovery at optical wavelengths,[2] which went on to produce pioneering work in the study of the surface structure and atmospheric stratification of red giants and supergiants.[3–7] A far wider realm of astrophysical targets was opened by the first major program on a 10 m class telescope: the Keck masking experiment.[8] Operating in the near-infrared and with snapshot two-dimensional Fourier mapping capability, this experiment revealed elaborate halos surrounding dying stars,[9–13] the first images of self-luminous disks circling young stars,[14–16] and the spectacular plumes around dusty Wolf-Rayets.[17–21]

(Send correspondence to P.T. E-mail p.tuthill@physics.usyd.edu.au)

Historically, the penalty for the improvement in image fidelity has been the restriction of the technique to very bright targets. Masks typically pass from a few to ten percent of the incident starlight, and exposures were very short to freeze the phase evolution of the atmosphere. More recently, masking has been performed downstream of adaptive optics (AO) systems with this hybrid "sparse aperture AO" proving to be capable of leveraging the advantages of both of its parent technologies. The Sparse Aperture Masking (SAM) mode offers a solution to the perennial problem of PSF characterization in AO imaging: synthesizing the heritage of image reconstruction from astronomical interferometry with the long coherence times available after AO correction. Marrying these two techniques therefore brings the precision calibration shown to come with sparse-aperture Fizeau-type interferometry together with extremely high dynamic range and the relatively faint target capability of AO.

Although SAM observations can deliver a unique and incisive datasets to a very wide variety of astronomical topics, among the most exciting and actively developed area in recent years has been the discovery of companions (such as brown dwarfs and planets) at very high contrast ratios. Here, the exquisite calibration properties of the masking interferometer enable the characterization of the point-spread-function of the instrument to very high precision.[22] Whereas orthodox Lyot coronagraph configurations are limited to searches beyond $4\,\lambda/D$ (where $\lambda$ is the wavelength and $D$ the telescope diameter) and only attain high contrast well beyond this, we have found that SAM is able to perform well in the region from several $\lambda/D$ down to separations as small as $0.5\,\lambda/D$. It turns out that this separation range is a particularly valuable one. When imaging the nearest populations of young stars, AO surveys to date can be criticized for doing little more than proving that planets are rare in wide orbits where they are not expected to form anyway. SAM on the other hand has been able to probe solar system scales, successfully revealing companions and brown dwarfs at resolutions consummate with the diffraction limited core.[23–26]

This report proceeds with an overview of the hardware and procedures associated with SAM mode at NAOS/CONICA. Following this, we describe results from commissioning various operational modes, including the imaging of complex targets, the detection of high contrast companions, and finally the implementation of an optical interferometric polarimetry mode utilizing the Wollaston prisms in the camera. We conclude with notes about the software support and data analysis pipeline which has been made available to workers using SAM mode.

## 2. THE NAOS/CONICA APERTURE MASKING SETUP

The CONICA camera was provisioned with an extensive set of internal wheels to carry filters and optical components into the beam. By designing masks for four different slots, SAM mode is able to perform for a far wider range of scientific targets than would be the case for a single mask geometry. Masks were fabricated by precision numerically-controlled laser milling of 0.2 mm steel stock. Parts installed into CONICA are depicted in Figure 1.

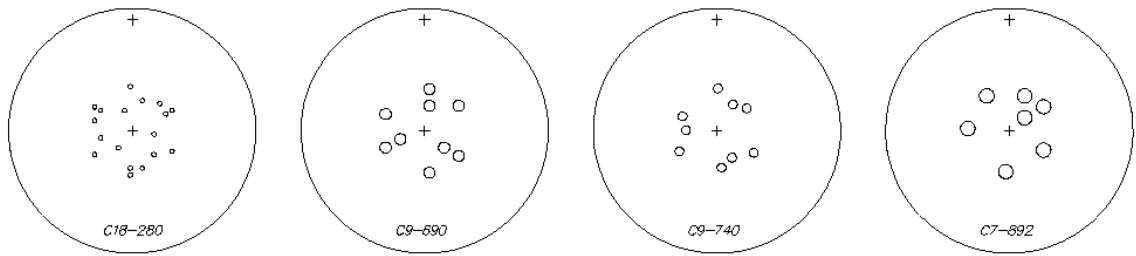

**Figure 1.** Laser-cut steel masks of 20 mm outer diameter with hole patterns (from left to right) of the 18holes, 9holes, BB-9holes and 7holes masks.

Masks can be inserted into the beam at a pupil plane by the camera software. When viewing a star, the effective primary mirror geometry produced by each mask is depicted in Figure 2. The different layouts are intended to broaden the range of astronomical targets accessible to the experiment. The three most commonly used masks are the "7holes", "9holes" and "18holes", which offer a sequence from a few large holes through to many small holes. The most basic difference between these masks is the fraction of starlight passed, which is 16%, 12.1% and 3.9% respectively. Masks which pass more light are suited to progressively fainter targets, however this is not the only consideration to be applied in formulating an observing strategy. With fewer holes, one also obtains less Fourier data and objects which require the mapping of complex structures become increasingly difficult. Therefore, the 7holes mask is really intended for studies of simple structures (such as binaries or companions) in the faintest class of targets. The 18holes on the other hand, is well suited to bright sources such as cool giant stars and dusty circumstellar environments. The intermediate 9holes has very good Fourier coverage and with a penalty of only 25% less throughput, should be used in preference to the 7holes for all but the very faintest of targets.

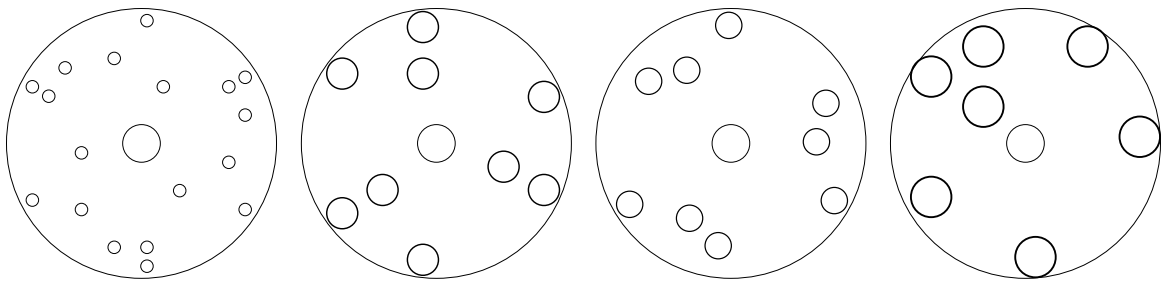

**Figure 2.** Masks installed into the pupil wheel of the camera apodize the telescope pupil into the desired sparse array geometry. When projected onto the 8 m primary mirror (large circles), the effective pupils of the four mask geometries are *from left to right:* "18holes" mask, "9holes" mask, "BB-9holes" mask and "7holes" mask.

A somewhat separate case not discussed so far is the BB-9holes mask, which is intended specifically for very broad bandwidths (hence "BB"). The Fourier coverage is designed to be resistant against *bandwidth smearing* in which the spatial frequency components are spectrally broadened by the wide bandpass to the extent that different baselines of the nominally non-redundant geometry begin to overlap in the $uv$ plane. When this happens, baselines are aliased together and the data become difficult to interpret. Practical tests with commissioning data have found that there is only a narrow niche for using this specific anti-aliasing mask. The normal "9holes" mask can tolerate fractional bandwidths well in excess of 10% before smearing becomes an issue, and should be fine for the entire CONICA filter set with the exception of the full J and H filters (fractional bandwidth $\sim 20\%$). Here, for chromatic sources, the "BB-9holes" should be used in preference to the "9holes".

Note that masks with small holes, such as the 18holes, can also be used only in combination with the narrowband filters (due to the bandwidth smearing discussed above). Furthermore, the small holes spread the light over a wider area on the chip. These effects combine to make the spread in source flux appropriate to each mask a *much* steeper function than one would expect based on the ratio of throughput alone.

## 3. SAM IN CONICA: PRACTICAL IMPLEMENTATION

Masks were designed to be exactly centered within the CONICA pupil wheel slots. Because it is desirable to explore the longest baselines available, the masks have holes near the mirror edge and only a 2% tolerance on misalignement of the pupil. The camera allows one degree of freedom in adjusting the mask position by rotation of the the wheel carrying the mask. After this adjustment, the best alignment for two masks is depicted in Figure 3.

As can be seen from the images, at the best alignment the pupil is significantly offset (by about 5%) from the centerline of the optical axis within the camera which results in the outermost holes towards the bottom of the pupil being vignetted. This clipping of the holes has a detrimental effect on the quality of the data recovered. Small changes in the optical path as the telescope and camera are moved around the sky result in variation of the amount of the light passing through the affected holes. This in turn compromises the calibration of the data. In the worst cases (for the 18holes mask) from one to several holes must be discarded completely from the analysis as adequate calibration is not possible. An intervention to rectify this optical alignment as been planned for some time but not yet effected at the time of writing.

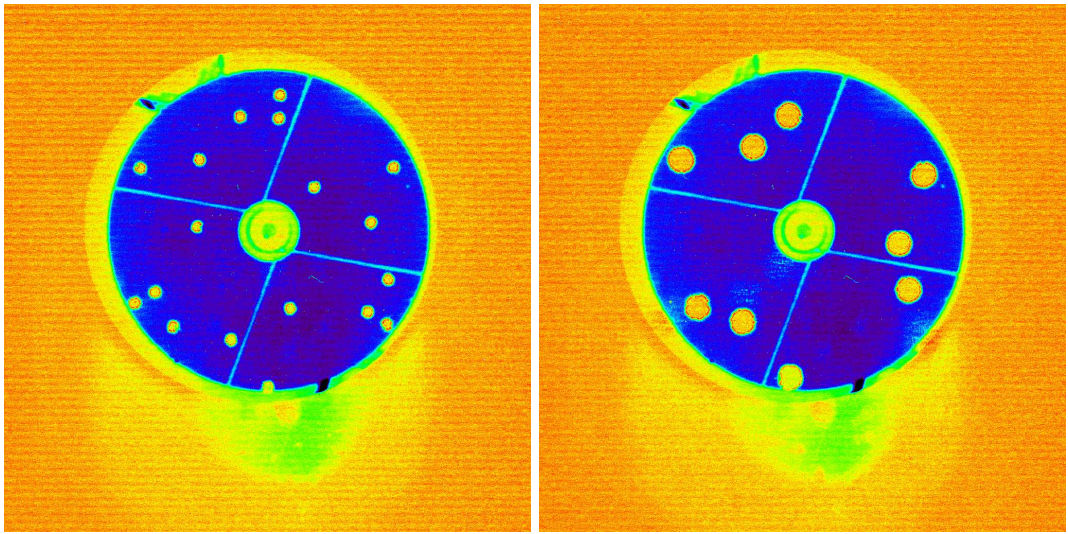

**Figure 3.** Images obtained with CONICA with the skylight-illuminated telescope pupil overlayed with the location of the mask holes for the 18holes mask *left* and for the BB-9holes *right*.

A second source of variable blocking of the mask holes will be immediately apparent from casual examination of Figure 3. In the normal course of operations, telescope spiders will rotate with respect to the camera as the sky's parallactic angle is tracked by the instrument. For masking (and other high fidelity techniques) the change in optical configuration caused by this is unacceptable. SAM mode therefore pioneered the use of a "pupil-tracking mode" at VLT in which the control software tracks the orientation of the pupil and keeps this fixed with respect to the camera, and consequently allowing the sky to rotate. As each mask is different, a specific rotation angle tailored to avoid spider-hole crossings is encoded in the software.

## 4. IMAGING COMPLEX ASTROPHYSICAL TARGETS

In order to explore the capability to recover full images at diffraction limited resolutions of relatively complex targets, the 18Holes mask was used. This gives the best Fourier coverage and well-sampled short and long baseline data, but restricts observations to relatively bright systems. An alternate mask geometry which has proved to be very successful at the imaging fainter complex targets is an annulus geometry: this forms one possible upgrade path now being explored with ESO staff.

### 4.1. VY Canis Majoris

VY Canis Majoris is a spectacular extreme M-supergiant which has produced an extensive infrared nebula several arcsec in extent. At the core, VY CMa exhibits a bright asymmetric plume, first imaged in detail by the Keck masking project.[27] Figure 4 shows these published images from data taken in 1997, together with new images produced in narrowband filters within the H and K bands using 18Holes mask with SAM at CONICA. Note that the asymmetric structure reproduced is nearly identical with that found at Keck over a decade ago. The high level of image fidelity in recovery of structure at angular scales up to and beyond the formal diffraction limit is the signature contribution of SAM mode to astronomical imaging.

For comparison, we also show the results of imaging observations using the full telescope pupil taken only minutes apart from the SAM data. With the use of the adaptive optics system, we have taken an identical series of rapid exposures to those obtained for the SAM observations, and used the shift-and-add algorithm to stack these data into a final resultant best image. This is given in Figure 5. There is some correspondence between the AO-only and SAM, in that there is evidence for a similarly skewed center of brightness in the AO image. However, the fine detail and diffraction-limited structures appearing in the masking data cannot be seen in the AO image. It is possible that with deconvolution using a carefully recorded PSF frame that more real structure may be recovered from the AO, but this procedure has proved to be controversial in the past, and can lead to spurious structures.

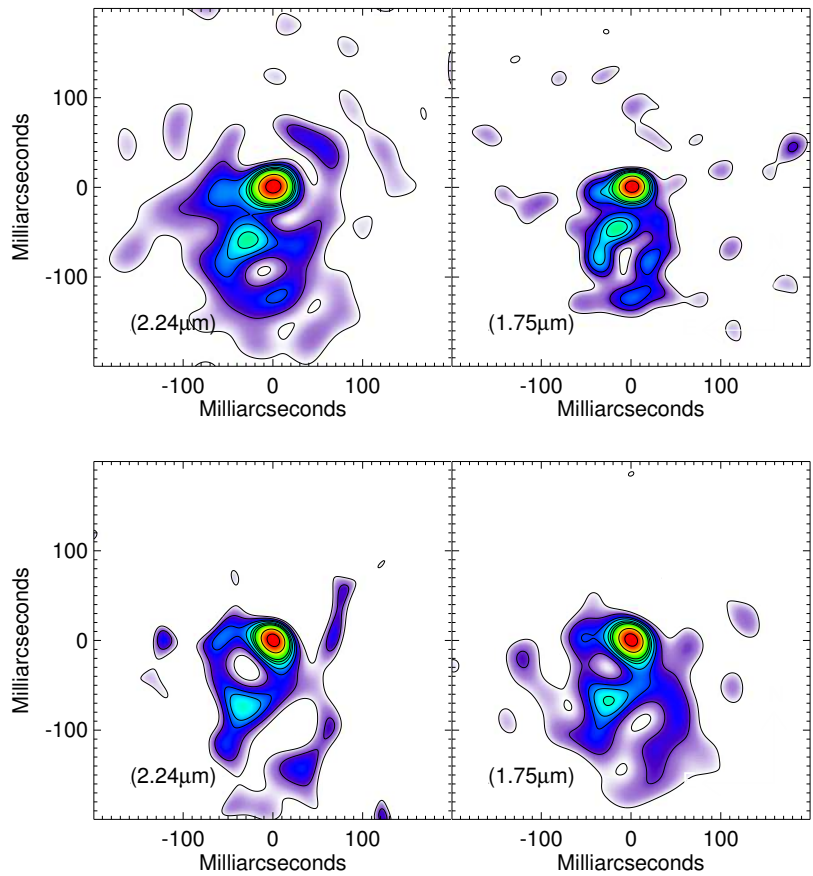

**Figure 4.** Images of the extreme mass-losing supergiant VY CMa. Upper panels show published maps recovered from the earlier (seeing-limited) aperture masking experiment from data taken January 1997.[27] Lower panels show images recovered with NACO/SAM data taken in March 2008. Not only is it possible to demonstrate recovery of all the earlier features, but apparent motion of the dust plume is apparent over the ∼11 year interval since the original images.

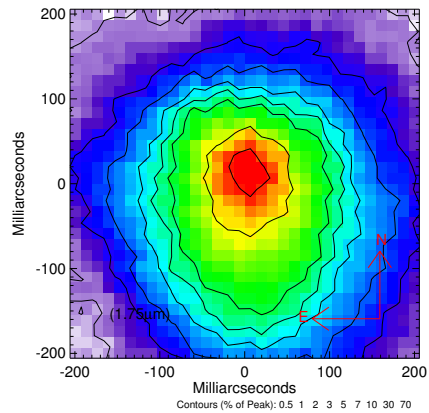

**Figure 5.** Supergiant VY CMa imaged in a more orthodox imaging mode with the full telescope pupil and adaptive optics system correction. NACO data taken in March 2008 with a narrowband K filter.

## 5. DETECTION OF COMPANIONS AT HIGH DYNAMIC RANGE

In recent years, masking interferometry has set the standard for the routine recovery of high contrast imaging data over scales from several down to $\sim 0.5\,\lambda/D$. In the discovery and characterization of orbits of brown dwarfs[23–25] together with ambitious surveys penetrating well into the planetary mass regime[28] the technique presently enjoys nearly uncontested access to the critical realms of imaging parameter space within a few resolution elements from the diffraction limited core.

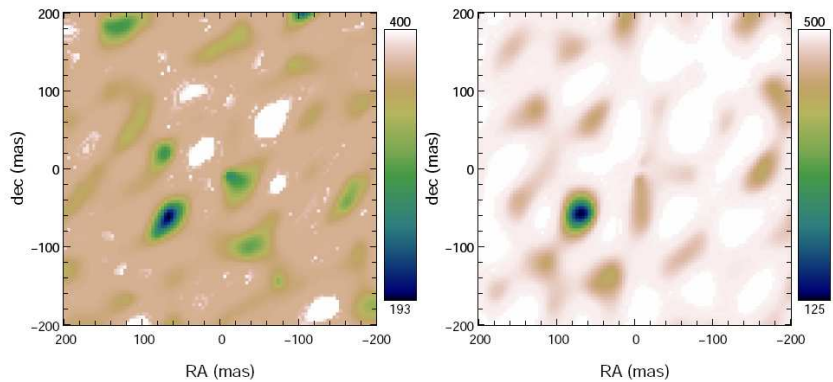

**Figure 6.** A chi-squared exploration of potential binary companion location from the vicinity of the star BD-21 4300. The *left panel* shows data at H-band, while the *right panel* shows K band. Note the strong minimum in Chi-Squared in both panels indicating the location of the companion.

Although visibility amplitudes recovered by masking interferometry do exhibit significant signal-to-noise gains over the filled-aperture, in practice the precision achieved is not sufficient to reveal small signals from companions fainter than contrasts of 10–20 (although performance varies greatly with conditions). For this reason, visibility amplitude data are usually discarded in the software and the companion detection process relies entirely upon the *closure phase* signals. Closure phases[2,29] are inherently self-calibrating, are not biased by the seeing, and they obey quasi-Gaussian statistics. Exploiting the technique of examining closure phase data for characteristic signals betraying the presence of a binary companion, contrast ratios in the realm of several hundred to one have been demonstrated[28] at scales at and somewhat beyond the diffraction limit.

### 5.1. High contrast companion to BD-21 4300

As an illustration of these methods of high contrast detection with SAM at CONICA, we show data from the system BD-21 4300 which has a known companion[28] at 4 mag contrast and 90 mas separation (note this is only 1.3 $\lambda/D$ for the 2.17 $\mu$m wavelength of observation). Closure phase signals on all closing triangles were recovered from SAM data obtained in March 2008 and explored for systematic signals over the multi-dimensional space of all possible separations and contrast ratios.

Figure 6 shows the resultant Chi-Squared parameter space for candidate detections within the vicinity of the host star out to separations of 200 mas. The clear minima here in both the H and K band data yield the best-fit separation and position angle for the binary companion. To get a feel for the degree to which the data really do exhibit this signal, the optimal phases have been reduced from the closure phase data and plotted against the baseline projected onto the best fitting binary orientation. A binary will now exhibit a systematic departure from random noise in the form of a sinusoidal signal, as seen clearly in Figure 7.

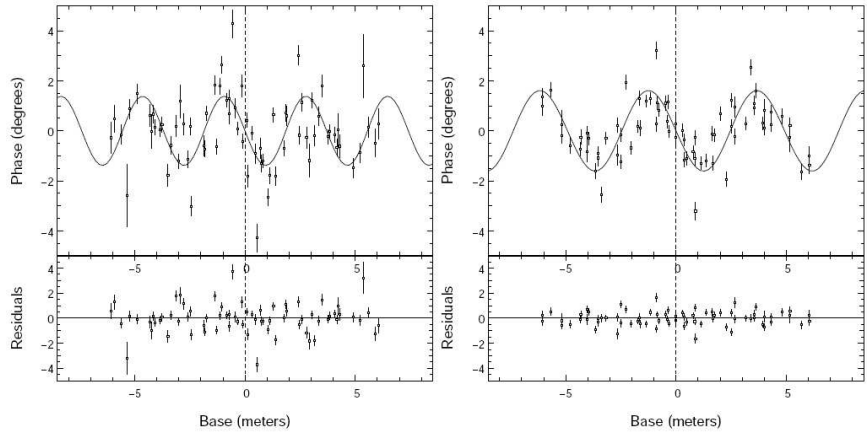

**Figure 7.** Best-fitting binary signal (sinusoidal line) to the phase data (points) obtained from SAM observations of BD-21 4300. The *right panel* shows data at H-band, while the *left panel* shows K band. On the horizontal axis is the baseline length projected onto the binary separation vector, with the vertical axis giving the magnitude of the phase signal in degrees.

## 6. OPTICAL INTERFEROMETRIC POLARIMETRY (SAMPOL)

In the optical/IR, celestial targets usually exhibit a polarized signal due to scattering or reflection of light. Common astronomical objects studied through polarimetry include young stellar objects, evolved mass-losing stars and solar-system bodies. Where such scattering occurs in the immediate vicinity of a star (as opposed to an extended nebula), it can often be the case that the polarized signature is very difficult to observe. This is because light from different spatial regions, each of which may be highly polarized, often adds together with polarization vectors which cancel out any net signal when the final stellar image is formed. The integrated polarization from even highly polarized stellar targets is therefore rarely more than a few percent, despite the fact that parts of their circumstellar environments probably emit nearly completely polarized light.

What is needed is an observing method capable of preserving these polarized signals which arise from regions which may be only milli-arcseconds apart on the sky. It is the aim of Optical Interferometric Polarimetry (OIP) to gives us such a technique, capable of delivering simultaneous polarimetry and high spatial resolution.

Utilizing SAM mode with the polarizing Wollaston prisms and half-wave plates already installed as part of the CONICA optics complement, we have commissioned a powerful new observing mode, christened SAMPol. It has been mentioned earlier in this report that calibrated visibility amplitude signals often have precisions of no better than 5–10%; this is not sufficient for revealing small polarization signals. Fortunately, it is possible to dramatically enhance the calibration precision of visibility measurements in polarized light by exploiting differential techniques. The half-wave plate can be used to exchange the polarization states of the two images produced by the Wollaston, allowing for a high degree of resilience against instrumental and seeing induced sources of measurement error of visibility amplitudes.

In order to test the ability of SAMPol to recover signals from high resolution polarized sources, the late-type giant W Hydrae was observed utilizing the differential interferometric polarimetry methods sketched above. After reduction of the raw data, the final observables can be obtained as the polarized Stokes Q and U parameters as a ratio against the total intensity Stokes I. Data plotting these quantities as a function of the azimuth of the baseline are given in Figure 8 below.

Identical plots produced for the PSF reference star (which exhibits no dust scattering and hence no polarization signal) showed data scattered randomly about zero with no systematic signals present. Similarly, data obtained by taking the ratio of visibilities in polarization states that had been rotated by a full 180 degrees also showed zero net signals, as expected from basic principles. These null signals obtained where expected lend confidence to the veracity of the astrophysical origin of the data depicted in Figures 8.

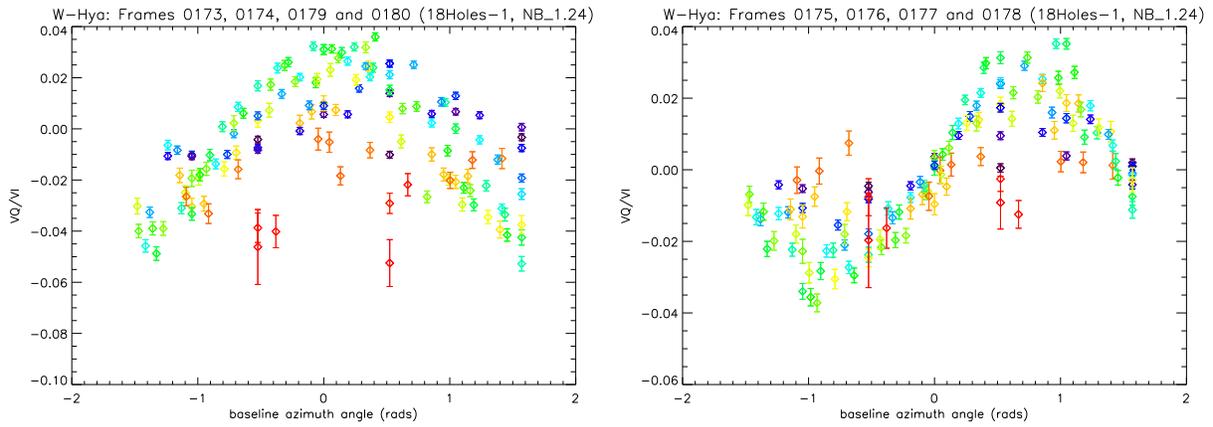

**Figure 8.** The processed $V_Q/V_I$ values for W Hya (instrumental Stokes coordinates) in J-band. The calibrated values of $V_I$ vary from 1.0 to ~0.1 at the longest baselines, as the target is resolved. Different baseline lengths are color coded; blue for short and red for the longest baselines passed by the aperture mask.

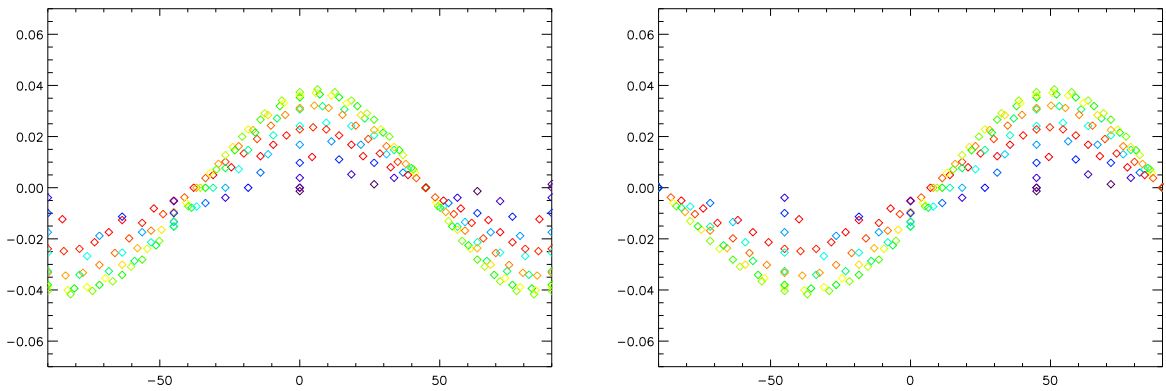

**Figure 9.** Expected $V_Q/V_I$ (Left Panel) and $V_U/V_I$ (Right Panel) signals from a toy model consisting of a thin polarized scattering disk surrounding a star with Uniform Disk diameter equivalent to W Hydrae. Functional form of signal from this model shows a close resemblance to the recorded data, and is indicative of a ~5% optical depth scattering from a dust shell at about 2 continuum radii from the surface of W Hya.

In order to explore the systematics of the data obtained, a simple toy model consisting of a uniformly illuminated circular disk with an apparent diameter appropriate to the literature size of W Hydrae was constructed. A totally polarized scattering ring was placed at a distance of about two radii above this disk, and the expected optical interferometric polarimetry signals generated. The results, which bear a strong resemblance to the recorded data, are depicted in Figure 9.

These results are highly significant, as observations uncovering polarization signals at similar spatial scales are extremely rare in the literature.[30] The commissioning of this mode is therefore demonstrating unique science reach for scattered-light polarized signals in previously unobtainable realms of angular scale.

# 7. DATA ANALYSIS AND SUPPORT

In commissioning this observing mode, extensive resources have been developed. These encompass observational planning, detailed optimization strategies for specific science outcomes, and post-observational data analysis support. An example of one of the observational planning tools available is the masking exposure-time calculator which runs under a graphical user interface in the IDL scientific data analysis language. A screenshot of the program is given in Figure 10, and has been made available by the Authors (email to P.T.) on request.

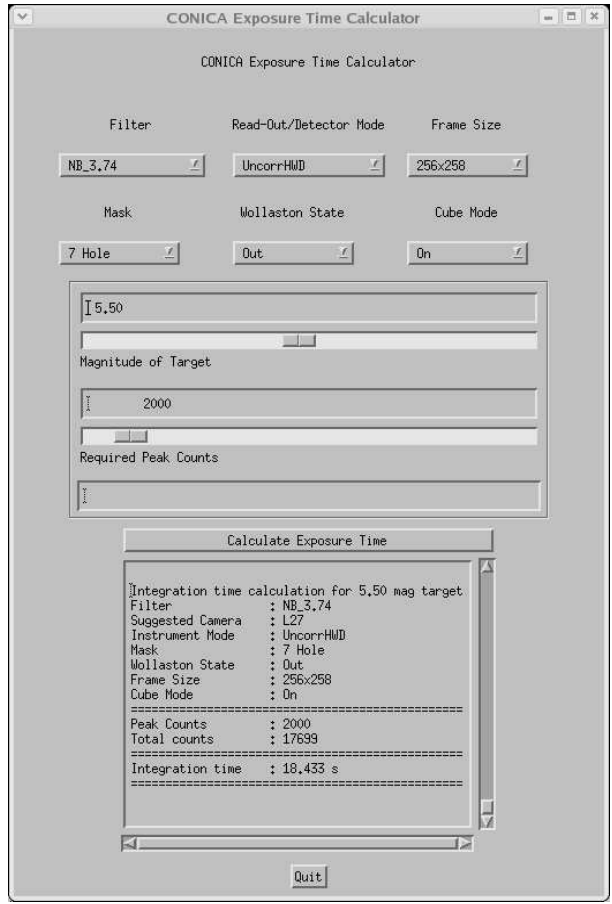

**Figure 10.** Graphical interface for the SAM exposure time calculator. This facilitates observational planning with simple sliders and pull-down menus to give recommended observing configurations based on a diverse range of settings of the camera optics and electronics.

At the VLT there has also been development of dedicated templates to implement the settings and configurations required to perform the experiment. Some of these, such as the "pupil-tracking mode" described earlier, were in themselves important additions to the capability of the observatory and have proved to be useful to a wider range of astronomical programs beyond the design specifically for SAM mode. Another of the operational mode developed for use specifically within the team is known as "Star Hopping" and facilitates the rapid jumping through a series of neighboring targets in the sky without incurring the large ($\sim$ 10 minute) penalty associated with the re-optimizing and locking of the adaptive optics control loops.

An extensive SAM data analysis suite based on Fourier transform techniques has been developed in IDL and released by the Sydney group. This is now freely available on a trial basis to the community, and can be obtained by contacting Peter Tuthill. A completely independent software pipeline has also been developed at Meudon by Sylvestre Lacour.

New discovery space has been opened on several fronts by the addition of the SAM mode at the CONICA camera. Of particular note are the high contrast imaging capabilities, and the recently-commissioned SAMPol mode of optical interferometric polarimetry. These have already led to unique new scientific discoveries. Future enhancements in the short term are likely to include the addition of an annulus mask to the suite of available geometries, enabling the imaging of a fainter class of complex targets.